\documentclass[twocolumn,showpacs,aps]{revtex4}
\usepackage[all]{xy}
\usepackage{amsmath}
\usepackage{amssymb}
\usepackage{graphicx}
\def\sc{\scriptscriptstyle}
\def\be{\begin{equation}}
\def\ee{\end{equation}}
\newcommand{\bes}{\begin{subequations}}
\newcommand{\ees}{\end{subequations}}
\newcommand{\wt}{\widetilde}
\def\bea{\begin{eqnarray}}
\def\eea{\end{eqnarray}}
\def\ba{\begin{array}}
\def\ea{\end{array}}
\def\ben{\begin{enumerate}}
\def\een{\end{enumerate}}

\begin{document}
\newcommand{\half}{{\textstyle\frac{1}{2}}}
\newcommand{\eqn}[1]{(\ref{#1})}
\allowdisplaybreaks[3]
\def\a{\alpha}
\def\b{\beta}
\def\g{\gamma}\def\G{\Gamma}
\def\d{\delta}\def\D{\Delta}
\def\ep{\epsilon}
\def\et{\eta}
\def\z{\zeta}
\def\t{\theta}\def\T{\Theta}
\def\l{\lambda}\def\L{\Lambda}
\def\m{\mu}
\def\f{\phi}\def\F{\Phi}
\def\n{\nu}
\def\p{\psi}\def\P{\Psi}
\def\r{\rho}
\def\s{\sigma}\def\S{\Sigma}
\def\ta{\tau}
\def\x{\chi}
\def\o{\omega}\def\O{\Omega}
\def\k{\kappa}
\def\pa {\partial}
\def\ov{\over}
\def\br{\\}
\def\ud{\underline}
\title{Traveling wave solutions of nonlinear partial differential equations}

\author{D. Bazeia$^a$, Ashok Das$^{b,c}$, L. Losano$^a$, and M.J. Santos$^{a,d}$}
\affiliation{$^a$Departamento de F\'{\i}sica, Universidade Federal da Para\'{\i}ba, 58051-970 Jo\~ao Pessoa PB, Brazil}
\affiliation{$^b$Department of Physics and Astronomy, University of Rochester, Rochester, NY 14627-0171}
\affiliation{$^c$Saha Institute of Nuclear Physics, 1/AF Bidhannagar, Calcutta 700064, India}
\affiliation{$^d$Centro Federal de Educa\c c\~ao Tecnol\'ogica de Sergipe, 49400-000 Lagarto SE, Brazil}

\begin{abstract}
We propose a simple algebraic method for generating classes of traveling wave solutions for a variety of partial differential equations of current interest in nonlinear science. This procedure applies equally well to equations which may or may not be integrable. We illustrate the method with two distinct classes of models, one with solutions including compactons in a class of models inspired by the Rosenau-Hyman, Rosenau-Pikovsky and Rosenau-Hyman-Staley equations, and the other with solutions including peakons in a system which generalizes the Camassa-Holm, Degasperis-Procesi and Dullin-Gotwald-Holm equations. In both cases, we obtain new classes of solutions not studied before.
\end{abstract} 
\maketitle

Finding traveling wave solutions of nonlinear partial differential equations has been of great interest primarily within the context of integrable systems \cite{b1,b2,b3,b4}. Such studies have led to many interesting types of solutions in the past such as the soliton solutions, the cnoidal solutions, the compacton solutions, the peakon solutions. However, finding these solutions has not been easy at all as is evidenced in the literature. 

In a recent paper \cite{bdl}, we proposed a simple method for generating traveling wave solutions of general nonlinear equations, which may or may not be integrable, starting from solutions of simple equations (including even linear equations). As we have demonstrated there through nontrivial examples (of mostly third order equations), this method is very powerful in obtaining traveling wave solutions of nonlinear equations. However, we understand now that the underlying reason behind the simplicity of our earlier proposal \cite{bdl} is some very remarkable properties of traveling wave solutions which allows us to propose even a simpler and more general method for generating classes of traveling wave solutions of a large class of partial differential equations starting from a trial traveling wave and an invertible map, which is the primary result of this letter. The earlier proposal \cite{bdl} forms only a part of this general result.

To explain the method, let us consider a (one) space and time dependent nonlinear $N$th order equation of the form 
\begin{equation}
{\rm F}(v,v_{t}, v_{x}, v_{2x},\cdots )=0,\label{eqn}
\end{equation} 
where $v(x,t)$ denotes the dynamical variable and the subscripts denote partial derivatives with respect to the corresponding variables
\begin{eqnarray}
v_{t}\!=\frac{\partial v}{\partial t},\;\; v_{x}\!=\frac{\partial v}{\partial x},\;\;
v_{nx}\!=\frac{\partial^{n}v}{\partial x^{n}},\;\; v_{nt,mx}\!=\frac{\partial^{n+m}v}{\partial t^{n}\partial x^{m}}.\label{dernotations}
\end{eqnarray}
We note that equation \eqref{eqn} may or may not be related to an integrable system. 
A traveling wave solution of the system has the form $v(x,t)=v(kx-\omega t)$, where $k$ denotes the wave number and $\omega$ the frequency of the traveling wave with the dispersion relation determined by the dynamical equation.

We note that for any given form of the traveling wave solution $v(kx-\omega t)$, all the $x$ and $t$ derivatives can be expressed in terms of the solution itself. Thus, for a traveling wave we can write 
\begin{equation}
v_{x}=V_{1} (v),\quad v_{t}=-({\omega}/{k})\,V_{1}(v),\label{firstorder}
\end{equation}
where the functional form of $V_{1} (v)$ can be explicitly determined if the form of $v$ is known. Let us denote $y= kx-\omega t$ and explain this assertion a bit more in detail. If $v = f (y)$ is defined by a (differentiable) bijective map (one-to-one and onto)  then $f$ has an inverse and $y= f^{-1} (v)$. In this case, we can write 
\begin{equation}
\frac{dv}{dy} = v_{y} =  f' (f^{-1} (v)) = g (v).
\end{equation} 
However, as we know many maps ($f$) in physics are not invertible \cite{inverse}. If $f$ is not a bijection (so that its inverse does not exist),  then we can still define an inverse of $f$ on each domain $D_i$ where the function $f(y)$ is monotonic (so that an inverse $f^{-1}_i$ exists on the domain $D_i$). In this case we can write 
\begin{equation}
\frac{dv}{dy}=f'(f^{-1}_i(v))=g_i(v),\ {\rm for}\ y \in D_i,
\end{equation}
where $f^{-1}_i$ is the inverse of $f$ on the domain $D_i$. The assertion in \eqref{firstorder} makes this assumption. Furthermore, if we make the identification $v\equiv V_0$,  then any higher derivative of $v$ can be written as 
\begin{equation}
v_{nx} =V_{n},\quad v_{nt,mx}=\left(-\frac{\omega}{k}\right)^{n}V_{n+m},\label{highervderivatives}
\end{equation}
where for continuous functions, $V_{n} (v)$ are related to $V_{1} (v)$ recursively as
\begin{equation}
V_{n} (v) = \frac{dV_{n-1} (v)}{dv}\, V_{1} (v),\quad n\geq 1.\label{vrecursion}
\end{equation}
For example, we note that
\begin{equation}
V_{2} = v_{xx}  = \frac{dv_{x}}{dx} = \frac{dV_{1} (v)}{dx} = \frac{dV_{1} (v)}{dv}\,v_{x} = \frac{dV_{1} (v)}{dv}\,V_{1} (v),
\end{equation}
and as a result, the expression \eqref{vrecursion} can be used recursively to express $V_n$ as a function of  $v$ when the form of $V_{1} (v)$ is known. 

Therefore, we see that for a traveling wave solution $v(kx-\omega t)$, the dynamical equation  \eqref{eqn}  of order $N$ reduces to an algebraic equation of the form 
\begin{equation}
{\overline{\rm F}}(\omega,k;v,V_{1},V_{2},\cdots ,V_{N})=0.\label{algeqn}
\end{equation}
This equation leads to relations between $\omega, k$ and the functions $V_{n} (v)$ and can also determine the dispersion relation for a traveling wave solution. The algebraic equation \eqref{algeqn} can be the starting point to check if a given ansatz $v (kx-\omega t)$ solves \eqref{eqn}. Namely, with a given ansatz for $v$, the form of $V_{1} (v)$ and, therefore, all the $V_{n}(v)$ would be determined and \eqref{algeqn} would reduce to a polynomial algebraic equation in $v$ which is undoubtedly much easier to analyze. However, our goal is to generate classes of traveling wave solutions with parameters which for different values lead to different kinds of solutions and we propose to do this in the following manner.

The method basically involves two steps. First, we choose a simple traveling wave $u (kx - \omega t)$ which we call a trial wave. Then we introduce an invertible map \cite{bdl,blm} 
\begin{equation}
u (kx-\omega t) = {\cal F} ( v (kx-\omega t)),\label{map}
\end{equation}
relating the trial wave to the kind of traveling wave solution we are looking for. It is worth noting here that this procedure may seem rather arbitrary. However, once a trial function is chosen (trigonometric, hyperbolic, $\cdots$) which is simple to manipulate and we keep in mind the kind of solution (soliton, compacton, peakon, $\cdots$) of the dynamical equation that we are interested in, the map can be constructed systematically with little arbitrariness (as we will describe in the examples below). The map is really chosen keeping the nature of the solution of interest in mind and  may involve several parameters. Requiring $v(kx-\omega t)$ to satisfy the dynamical equation \eqref{eqn} or equivalently \eqref{algeqn} may determine some of these parameters while the undetermined parameters would lead to a class of solutions depending on these parameters.

Checking if $v (kx-\omega t)$ given by the inverse of the map in \eqref{map} satisfies the dynamical equation is quite easy in its algebraic form \eqref{algeqn}. For example, the trial wave will lead to a relation of the form $u_{x}=U_{1}(u)$ (see \eqref{firstorder}), where $U_{1}(u)$ is known explicitly from the known form of the trial wave and  the map \eqref{map} leads to 
\begin{equation}
V_{1} (v) = \frac{U_{1}({\cal F} (v))}{{\cal F}^{\prime} (v)},\quad {\cal F}^{\prime} (v) = \frac{\mathrm{d}{\cal F} (v)}{\mathrm{d}v}.\label{V1map}
\end{equation}
Namely, the known from of the trial wave  $u$ as well as the choice of the map ${\cal F}$ would determine the (polynomial) forms of all the unknown functions $V_{n}(v)$ upon using \eqref{vrecursion}. As a result, the dynamical equation \eqref{algeqn} will reduce to an algebraic (polynomial) equation in $v$. The solution of this simple algebraic equation would possibly determine some (or all) of the parameters in the map and would generate a traveling wave solution of the original dynamical system \eqref{eqn} through the inverse map (see \eqref{map}) $v(kx-\omega t)={\cal F}^{-1}(u(kx-\omega t)).$ If all the parameters of the map are determined by the equation, then we have a particular traveling solution of the system (as in \cite{bdl}). However, if the map is chosen to be rich enough and some of the parameters are left undetermined by the equation, then this can lead to interesting classes of traveling wave solutions as we will illustrate through examples.

It is clear that this method applies equally well to equations which may or may not be integrable, and so the potential for its applicability is indeed quite great. However, we point out here that if there are discontinuities in the solutions (as some of the interesting solutions do), then the simple recursion relation \eqref{vrecursion} may not hold and a direct evaluation of the higher derivatives may become  necessary. This would be true in general when the derivative $v_x$ (or $u_{x}$), which leads to $V_1$ (or $U_{1}$), is not a single-valued function of $v$ (or $u$). On the other hand, if the dynamical equation \eqref{eqn} does not involve $v_x$ directly, rather $f(v_x)$ which is free from ambiguity, then our method can be carried out easily as we explicitly show in the examples below.

Let us explain the method as well as its potential with a couple of examples. We first study traveling waves of the form of compactons which are configurations of finite extent free of exponential tails and which arise as solutions of the Rosenau-Hyman (RH) equation \cite{RH}. Here we would study the deformed equation
\be\label{newa}
v_t+(P(v))_x+ (Q(v) v_{xx})_x=0
\ee
where $P(v)=\alpha_1 v+\alpha_2v^2+\alpha_3\,v^{2+p}+ v^{2+2p},$ and $Q(v)=v^{1+2p},$ with $\alpha_1,\alpha_2,\alpha_3,$ and $p>0$ real constant parameters. Equation \eqref{newa} is motivated by the RH family of equations $v_t+(v^m)_x+(v^n)_{xxx}=0$ \cite{RH}, as well as by the $N$ dimensional Rosenau-Pikovski (RP) \cite{RP} and the Rosenau-Hyman-Staley (RHS) \cite{RHS} family of equations $v_t+(v^m)_x+(1/s)(v^r\mbox{\boldmath$\nabla$}^2_{\sc N} v^s)_x=0$, where a unidirectional convection is balanced by a $N$-dimensional dispersive force. Although the deformed model \eqref{newa} is one dimensional, we will show below that the nonlinear convection and dispersion which are present in \eqref{newa} bring a variety of effects as we change the parameters. Since the results are analytical, our procedure may be of some use in the search of analytic solutions in higher dimensions, an issue which is beyond the scope of the present work. For a traveling wave solution of \eqref{newa}, we proceed as discussed earlier and obtain the relation (see \eqref{algeqn}) 
\be
\frac{\omega}{k}-P^\prime(v)-Q^\prime(v)V_2(v)-Q(v)V_2^\prime(v)=0,\label{RHreducedeqn}
\ee
where a prime denotes derivative with respect to the argument as in \eqref{V1map}.

To study compactons, let us choose the trial wave in the form $u(kx-wt)=\cos(kx-wt)$, for $|kx-wt|\leq\pi$, and $-1$ otherwise, which leads to $u_x=U_1(u)=-k\,{\rm sgn}(kx-wt)\,\sqrt{1-u^2}$ for $|kx-wt|\leq\pi$. The presence of the ${\rm sgn}$ function seems to introduce an ambiguity, but we note that  \eqref{RHreducedeqn} only depends on $V_2$ and its derivative so that there is no ambiguity in the equation. 

Let us next introduce the map $u=a+v^{p},$ where $a, p$ are real constant parameters so that we can write $v(kx-\omega t)=(\cos(kx-\omega t)-a)^{1/p}$. In this case, for $|kx-wt|\leq\pi$ the trial wave and its derivatives are continuous and we can follow our general procedure. Thus, we use  \eqref{vrecursion} and \eqref{V1map} to determine 
\begin{equation}
V_{2}(v)\!=\!\frac{k^{2}}{p^{2}}\!\!\left(\!-v\!+\!a(p\!-\!2) v^{1-p}\!\!+\!(1\!-\!p)(1\!-\!a^{2})v^{1-2p}\right)\!.
\end{equation}
Using this in \eqref{RHreducedeqn} we obtain the algebraic equation
\bea
&&\left(\alpha_1-\frac{\omega}{k}\right)+2\left(\!\alpha_2-\frac{k^2}{p^2}(1-p)(a^2-1)\right)v\nonumber
\\
&&+(2+p)\left(\alpha_3-\frac{k^2}{p^2}(2-p)a\right)v^{1+p}\nonumber
\\
&&+2(1+p)\left(1-\frac{k^2}{p^2}\right)v^{1+2p}=0.
\eea
This determines the parameters as $k = \pm p$, $\omega=\pm\alpha_1p$, $\alpha_2=(1-p)(a^2-1)$, and $\alpha_3=(2-p)a$.

We can now construct a variety of solutions with different choices of the above parameters. Let us also note here that since a constant is a solution of \eqref{newa}, by adjoining such a solution to the solution found with the inverse map in a continuous manner, one can easily construct new solutions of the dynamical system. For example, with $p$ an odd integer or the inverse of an odd integer, we obtain the compacton solution  
\be\label{R}
v\!=\left\{\begin{array}{ll}\left(\cos p(x\!-\!\alpha_{1} t)-a\right)^{\frac{1}{p}}\!\!,\;{\rm for}\;|p(x-\alpha_{1} t)|\leq{\pi},
\\-(1+a)^{\frac{1}{p}},\;\;\; {\rm otherwise,}\end{array}\right.
\ee
which satisfies \eqref{newa}. Note that if we redefine $v\to v+b$ in \eqref{newa}, this new parameter $b$ can be used to control the asymptotic behavior of the solution, such that for $b=(1+a)^{1/p}$ the shifted solution will vanish outside the interval $[-\pi,\pi]$.

\begin{figure}[ht!]
\includegraphics[width=3.2cm]{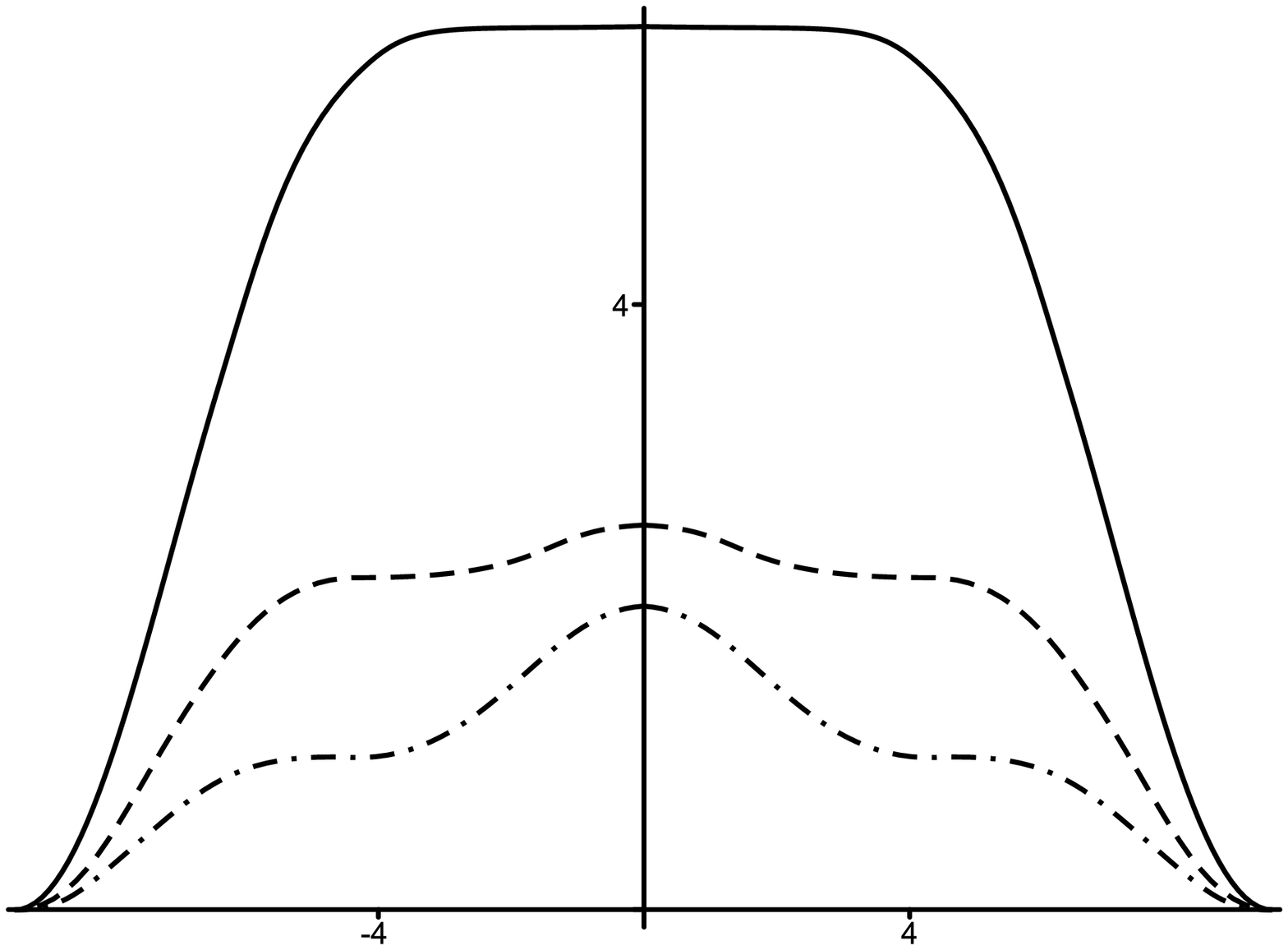}
\hspace{0.2cm}
\includegraphics[width=3.2cm]{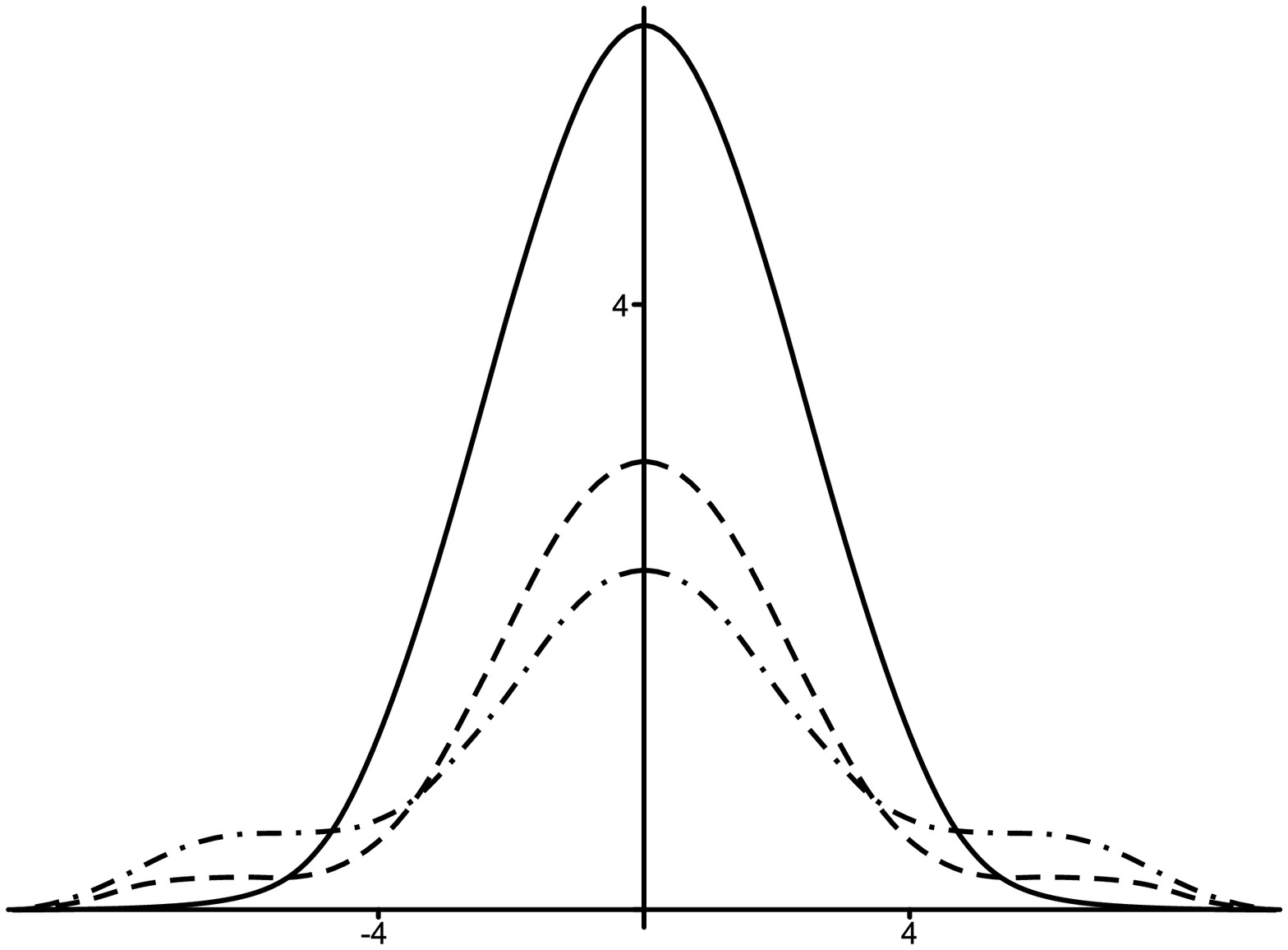}
\caption{{\small {Plots of the traveling wave solution for $p=1/3,$ with $a=0$, $a=0.3$ and $a=0.8$ (left panel), and with $a=-0.2$, $a=-0.4$ and $a=-0.8$ (right panel).}}}
\end{figure}

There is a variety of traveling wave solutions contained in \eqref{R} and some of them lead to interesting features which we now illustrate. In Fig.~1 we have plotted the solution \eqref{R} at $t=0$ for $p=1/3$, and for some values of $a\in[-1,1]$, where the amplitude increases with increasing $|a|$. We note that in the limit $a\to-1^+$ we have a compacton of standard bell shape. As $a$ increases, the solution develops a lump at the top which is a nice novelty for compactons, with the interesting feature that the height of the bottom $(h_b)$ and top $(h_t)$ portions of the new compactons obey: $h_b<h_t$ for $a\in(-1,0)$, $h_b=h_t$ at $a=0$, and $h_b>h_t$ for $a\in(0,1)$. We further note that in the limit $a\to1^-$, the solution leads to a compacton with a flat plateau.

In Fig.~2 we have plotted solutions at $t=0$ for $p=3,$ and for some values of $a,$ and there the amplitude increases with decreasing $|a|$. These plots unveil
remarkable features of the traveling waves: for $a=\pm0.5$ the solution is a [thin $(+)$ or thick $(-)$] tipon-like compact wave \cite{RP}, for $a=1$ it is a peakon-like compact wave \cite{CH} (see below), and for $a=-1$ it has an oval-like compact shape. All of the above solutions identify distinct types of compactons,
some of them have never been studied before. In both cases, we note that the constant $a$, introduced by the chosen map, works like a deformation parameter, since it deforms the shape of the solution \cite{bdl,blm}.

\begin{figure}[ht!]
\includegraphics[width=3.2cm]{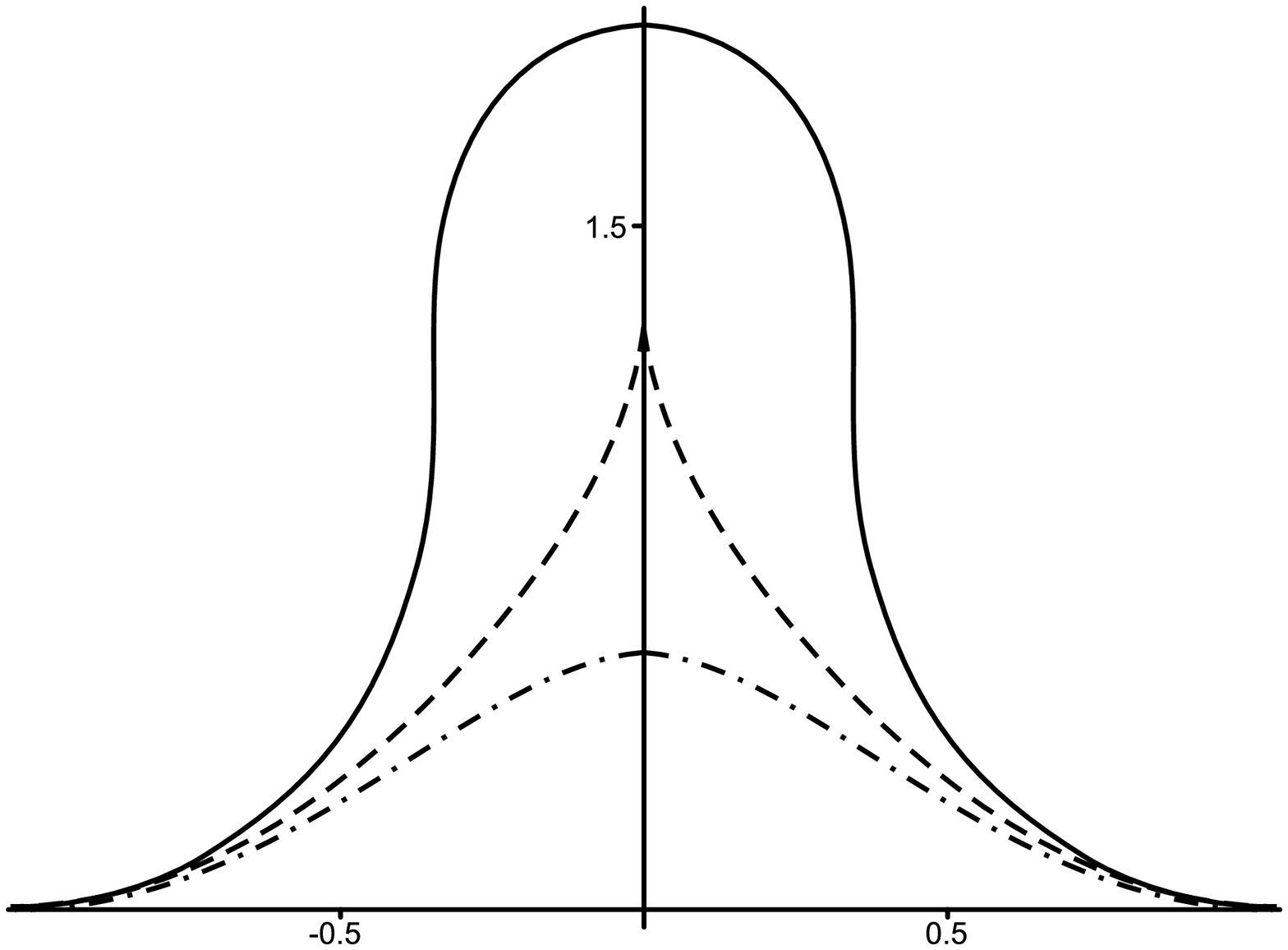}
\hspace{0.2cm}
\includegraphics[width=3.2cm]{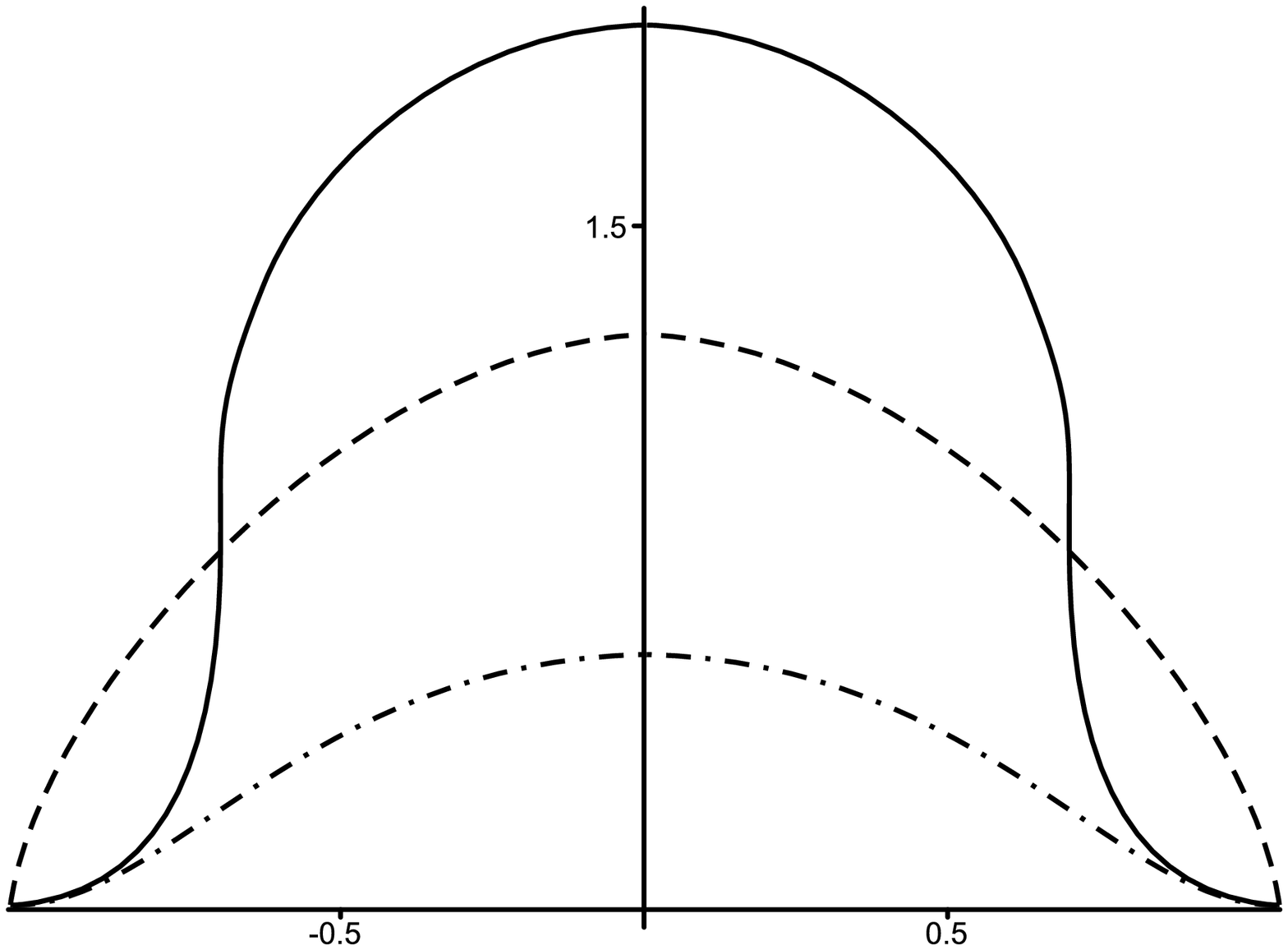}
\caption{{\small{Plots of the traveling wave solution for $p=3$ and for $a=0.5,$ $a=1$ and $a=1.5$ (left panel), and for $a=-0.5$, $a=-1$ and $a=-1.5$ (right panel).}}}
\end{figure}

To introduce the second example, let us recall that the equation $v_{t}-v_{xxt}+(r+1)vv_{x}-r v_{x}v_{xx}-vv_{xxx}=0$ is known to be integrable for $r=2,3$. For $r=2$ it is known as the Camassa-Holm (CH) equation \cite{CH} which was derived as an approximation to Euler's incompressible fluid motion. For $r=3$, it 
describes the Degasperis-Procesi (DP) equation \cite{DP} and both these equations represent models of shallow water waves \cite{b4}. Furthermore, both these equations are known to possess peakon solutions which have discontinuous first order derivatives. Since solitary waves constitute a well known form of the water waves, another integrable equation was recently introduced by Dullin-Gattwald-Holm (DGH) \cite{DGH}, which contains the KdV equation as well as the CH equation in different limits, thereby  having the interesting feature of supporting both the bell shaped solution of KdV as well as the peakon solution of CH as limiting cases.

Keeping this in mind, let us study, as another example of our method, a deformed system of the form
\begin{equation}
v_{t}-v_{xxt}+f(v) v_{x}-g(v) v_{x}v_{xx}-h(v) v_{xxx}=0,\label{newb}
\end{equation}
with
\begin{eqnarray}
&& f (v)=\beta_{0}+\beta_{1}v+\beta_{2} v^{2},\nonumber\\
&&g(v)=r+s v,\;\;\;\;\;h(v)={\wt\beta}_0 +{\wt\beta}_{1}v+{\wt\beta_2}v^2.\label{deformation}
\end{eqnarray}
We note that when $\beta_{0}={\wt\beta}_0=\beta_{2}={\wt\beta}_2=s=0, \beta_{1}=(r+1)$ and ${\wt\beta}_1=1$, equation \eqref{newb} reduces to the CH-DP equations (for $r=2,3$) while for $\beta_2={\wt\beta}_2=s=0$ it leads to the DGH equation \cite{DGH}. For general $\beta_{i}, {\wt\beta}_i, i=0,1,2$ and $r,s$ the deformed equation \eqref{newb} is indeed a much richer system and to clarify our method we try to find the traveling wave solution of this generalized system. 

A traveling wave solution of \eqref{newb} of the form $v(kx-\omega t)$, would satisfy a relation of the form (see \eqref{algeqn})
\begin{eqnarray}
&&\frac{\omega}{k}-\beta_{0}-\beta_{1}v-\beta_{2}v^{2}+(r+sv)V_{2}\nonumber\\
&&\quad+\left({\wt\beta}_0+{\wt\beta}_1 v+{\wt\beta}_2 v^2-\frac{\omega}{k}\right) V_{2}^{\prime}= 0.\label{CHreducedeqn}
\end{eqnarray} 
Keeping a soliton solution in mind, let us next choose a simple trial wave of the form $u(kx-\omega t)={\rm cosh}(kx-\omega t)$. Here we note that only $V_2$ and its derivative are present in \eqref{CHreducedeqn} so that the apparent ambiguity with the sign in $V_{1}$ is again not present. With this, we choose a map \eqref{map} of the form (keeping the soliton in mind) 
$u=\sqrt{{a}/{v}}$, where $a$ is a constant parameter to be determined. In this case, following our earlier discussion, we obtain
\begin{eqnarray}
V_{2}= 4k^{2}v - 6k^2v^{2}/a.
\label{VnCH}
\end{eqnarray}

As a result, \eqref{CHreducedeqn} leads to the algebraic equation for $v$ of the form
\begin{eqnarray}
&&\lefteqn{\left(\left(\frac{\omega}{k}-\beta_0\right)-4\left(\frac{\omega}{k}\!-\!{\wt\beta}_0\right)k^2\right)}\nonumber\\
&&+\left(12\left(\frac{\omega}{k}\!-\!{\wt\beta}_0\right)\!\frac{k^2}{a}\!+\!
4(r\!+\!\wt{\beta}_1)k^{2}\!-\!\beta_{1}\right)v\label{CHalgebraiceqn}\\
&&-\left(\!\!\beta_{2}\!+\!\left(6(r\!\!+\!\!2\wt{\beta}_1)\!\!-\!\!4a(s\!\!+\!\!\wt{\beta}_2)\right)\frac{k^{2}}{a}
\!\right)\!\!v^{2}\!\!-\!6(s\!\!+\!\!2\wt{\beta}_2\!)\frac{k^2}{a}\!v^3\!\!=\!0.\nonumber
\end{eqnarray}
Requiring the coefficients of each power of $v$ to vanish now determines the parameters as well as the traveling wave solution for \eqref{newb}.

There are several solutions of \eqref{CHalgebraiceqn} depending on the values of the parameters in \eqref{newb}. If ${\wt\beta}_0=\beta_0, {\wt\beta}_2=s=0$ and $\beta_2 = r+{\wt\beta}_1\neq0$, then \eqref{CHalgebraiceqn} determines $k=\pm1/2$ and
\begin{equation}
\omega\!=\!\pm\frac{(\beta_2\!+\!{\wt\beta}_1)(\beta_2\!-\!\beta_{1})}{4\beta_{2}}\!\pm\!\frac{\beta_0}{2},\;\; a\!=-\frac{3(\beta_2+{\wt\beta}_1)}{2\beta_{2}},\label{CH1stsoln}
\end{equation}
and the traveling wave solution corresponds to
\begin{equation}
v\!=\!\!-\frac{3(\beta_2\!+\!{\wt\beta}_1)}{2\beta_{2}} {\rm sech}^{2}\!\!\left(\!\frac{x}{2}\! \mp\! 
\frac{((\beta_2\!+\!{\wt\beta}_1)(\beta_2\!-\!\beta_{1}\!)\!+\!2\beta_0\beta_2)t}{4\beta_{2}}\!\right)\!\!.\label{CHbellsoln1}
\end{equation}
We note from this that there are several distinct choices of parameters which lead to waves of both positive and negative velocities. Furthermore, in our approach, $\beta_{2}$ is merely a deformation parameter and if we choose $\beta_{2}<0$, we indeed have a bell shaped solution.

On the other hand, for ${\wt\beta}_2=s=0, \omega = k\beta_{0}=k {\wt\beta}_{0}$ we have 
\be
k=\pm\frac12\sqrt{\frac{\beta_1}{r+{\wt\beta}_1}}, \;\;\;a=-\frac{3(r+2{\wt\beta}_1)\beta_1}{2(r+{\wt\beta}_1)\beta_2},
\ee
with the traveling wave solution
\be
v=-\frac{3(r+2{\wt\beta}_1)\beta_1}{2(r+{\wt\beta}_1)\beta_2}\,{\rm sech}^2\frac12\sqrt{\frac{\beta_1}{r+{\wt\beta}_1}}(x-\beta_0 t).
\ee
We note that the velocity in this case is determined by the constant $\beta_0$ alone. There are other choices of parameters, leading to distinct bell shaped solutions, but we now focus attention on the peakon solution.

To obtain the peakon solution for the deformed equation \eqref{newb}, let us choose the trial wave to have the form
$u(kx-\omega t)=\exp(-|kx-\omega t|)$, which leads to $u_x=U_1=-k\,{\rm sgn}(kx-wt)\,u$. We note that the solution has discontinuous derivatives and as mentioned earlier, the recursion relation \eqref{vrecursion} does not hold. In this case, we need to calculate the higher functions explicitly.  Let us rewrite \eqref{newb} in the form
\be
\left(\left(f(v)-\frac{\omega}{k}\right)-g(v)V_2\right)V_1=\left(h(v)-\frac{\omega}{k}\right)V_3,
\ee
and note that the presence of $V_1$ in the left hand side and of $V_3$ in the right hand of this equation removes any ambiguity with the {\rm sgn} function. The next step is to use \eqref{V1map} and $u_{nx}=U_n(u)$ to calculate 
\bes\bea
V_2(v)&=&U_2({\cal F}(v))/{{\cal F}^\prime}-V_1^2{{\cal F}^{\prime\prime}}/{{\cal F}^\prime},
\\
V_3(v)&=&U_3({\cal F}(v))/{{\cal F}^\prime}-3V_1(V_2{{\cal F}^\prime}+V_1^2{{\cal F}^{\prime\prime}})^2{{\cal F}^{\prime\prime\prime}}/{{\cal F}^\prime}\nonumber
\\
&&+V_1^3(3{{\cal F}^{\prime\prime2}}-{{\cal F}^\prime}{{\cal F}^{\prime\prime\prime}})/{{\cal F}^{\prime2}}.
\eea\ees
The higher derivatives can also be calculated easily, but they are not necessary in the present case. We then choose the map \eqref{map} $u=(v-b)/a$, where $a$ and $b$ are real parameters. With this, after substituting the explicit forms of  $V_1,V_2,V_3$ we end up with the algebraic equation 
\begin{eqnarray}
\lefteqn{k\, {\rm sgn}(kx\!-\!\omega t)(v\!-\!b)\!\biggl(\!\left(\left(\frac{\omega}{k}\!-\!\beta_0\right)\!-\!\left(\frac{\omega}{k}\!-\!{\wt\beta}_0+rb\right)k^2\right)\!v}\nonumber\\
&\!+&\!\!\!\!((r+{\wt\beta}_1-sb)k^{2}-\beta_{1})v^{2} +((s+{\wt\beta}_2)k^2-\beta_2)v^{3}\!\biggr)\nonumber\\
&\!+&\!\!\!\! 2k^{3} a \delta^{\prime} (kx-\omega t)\left(\wt{\beta}_{0} + \wt{\beta}_{1}v+ \wt{\beta}_{2}v^{2} - \frac{\omega}{k}\right)=\!0.\label{peakonalgebraicrelation}
\end{eqnarray}
Here we have simplified some of the terms using the conventional relation ${\rm sgn}(kx-\omega t)\delta(kx-\omega t) = 0$. Also, using the identity
$k \delta^{\prime}(kx-\omega t)v^{n}=(\delta(kx-\omega t)v^{n})_{x}=k(a+b)^{n}\delta^{\prime}(kx-\omega t)$, the last term in \eqref{peakonalgebraicrelation}
can be simplified even further. 

There are distinct solutions of \eqref{peakonalgebraicrelation} depending on the values of the  parameters in \eqref{newb}, but the last term in \eqref{peakonalgebraicrelation} determines $\omega=k (\wt{\beta}_{0}+\wt{\beta}_{1}(a+b)+\wt{\beta}_{2}(a+b)^{2})$. We note that for $b=0$, and for $\beta_0={\wt\beta}_0, \beta_1=r+{\wt\beta}_1, \beta_2=s+{\wt\beta}_2$, equation \eqref{peakonalgebraicrelation} leads to $k=\pm1$ without any
restriction on $a$ and the traveling wave solution has the form
\begin{equation}
v (x,t)=a\, e^{-|x-\omega t|}.\label{vCH}
\end{equation}
If we further take the simple choice (among many) of the parameters, $\wt{\beta}_{0}=\wt{\beta}_{2}=0, \wt{\beta}_{1}=\pm 1$, we can identify $a=\omega$ and the solution \eqref{vCH} coincides in this case with the peakon solution in \cite{CH}.  However, a nice novelty of our solution \eqref{vCH} is that the amplitude does not need to be equal to the velocity anymore, and as a result, we can also have a peakon solution with positive amplitude traveling with positive or negative velocity. For $b\neq0$, there are peakons where $b$ can be used to control the width and the asymptotic behavior of the solution. All the above results show a rich diversity of solutions of \eqref{newb} some of which, to the best of our knowledge, have not been studied in the literature before.

In conclusion, in this letter we have proposed a simple method for constructing classes of traveling wave solutions of nonlinear partial differential equations. Exploiting the properties of traveling waves, the method converts the dynamical equation into a simple algebraic equation which is easy to analyze. The two steps of taking a simple trial wave and then choosing a map (with several parameters) between the trial wave and the unknown traveling wave solution lead in a very efficient way to classes of  traveling wave solutions. This method is a generalization of our earlier results, and we have demonstrated the versatility and the potential of our method by constructing the compacton solution for the Rosenau-Hyman equation where we have also shown the existence of newer kinds of compacton solutions, and the bell shaped solution as well as the peakon solution for the Camassa-Holm equation, bringing novelties for both the bell shaped and peakon-like solutions.
In particular, we have found some very interesting new  compact solutions, which are the tipon-, peakon- and oval-like compactons. The method will be further used to study other nontrivial examples of current interest in nonlinear science separately \cite{longer}.

The authors thank CNPQ, PRONEX-CNPq-FAPESQ, and the US DOE Grant number DE-FG 02-91ER40685.


\end{document}